\documentclass[pra,showpacs,twocolumn,nofootinbib]{revtex4}
\usepackage{graphicx}
\setkeys{Gin}{draft=false} 
\usepackage{bm,amssymb,amsmath}
\usepackage{dcolumn}
\usepackage{natbib}
\usepackage
{hyperref}

\newcommand{\Gspr}{\Gamma_\mathrm{spr}}
\def\veps{\varepsilon}

\begin{document}

\title{Electron recombination, photoionization and scattering via many-electron
compound resonances}
\author{V. A. Dzuba$^{1}$}
\author{V. V. Flambaum$^{1}$}
\author{G. F. Gribakin$^2$}
\author{C. Harabati$^{1}$}
\author{M. G. Kozlov$^{1,3,4}$}

\affiliation{$^1$School of Physics, The University of New South Wales,
Sydney NSW 2052, Australia}

\affiliation{$^2$School of Mathematics and Physics,
Queen's University, Belfast BT7 1NN, Northern Ireland, UK}


\affiliation{$^3$Petersburg Nuclear Physics Institute, Gatchina
188300, Russia}
\affiliation{$^4$St.~Petersburg Electrotechnical University
``LETI'', Prof. Popov Str. 5, 197376 St.~Petersburg}

\date{\today}

\begin{abstract}
Highly excited eigenstates of atoms and ions with  open $f$ shell are chaotic superpositions of thousands, or even millions of Hartree-Fock determinant states. The  interaction between dielectronic and multielectronic configurations leads to the broadening of dielectronic recombination resonances and relative enhancement of photon emission due to  opening of thousands  of radiative decay channels. The  radiative yield  is close to 100\%  for electron energy $\lesssim 1$ eV  and rapidly decreases for higher energies due to opening of many autoionization channels.  The same mechanism predicts suppression of photoionization and relative enhancement of the Raman scattering.  Results of our calculations of the recombination rate are in agreement with the experimental data for W$^{20+}$ and Au$^{25+}$.
\end{abstract}

 \maketitle

\section{Chaos and statistical theory}\label{sec:chaos}
It is well known that a long-time
behaviour of a classical chaotic system is unpredictable due to exponential
divergence of phase-space trajectories. Any small changes in the initial conditions or computer rounding errors are exponentially enhanced over
time. However, the chaos  makes statistical predictions possible. For example,
we cannot predict the motion of a specific molecule in a gas, but we can predict the diffusion coefficients, distribution of velocities, pressure, etc.

In isolated quantum many-body systems chaos emerges due to the exponential growth in the energy level density caused by the increase in the number of ``active'' particles. It follows the increase in the energy of the system,
which allows more particles to be excited into unoccupied orbitals.
Indeed, the number of ways to distribute $n$ fermions over $m$ orbitals is exponentially large for $n,m\gg1$, even when $n$ itself is not very large. 

By distributing the particles among orbitals in different ways one generates the Stater determinant states $|i\rangle$ (configuration states) from some mean-field, e.g., Hartree-Fock, single-particle orbitals. These states serve as the basis for finding the 
\textit{eigenstates} $|n\rangle = \sum_i C^{(n)}_i |i\rangle$.
When the residual interaction between the particles exceeds the energy spacing between the basis states coupled by this interaction, the eigenstates become chaotic superpositions of thousands or even millions of basis states $|i\rangle$.

The expansion coefficients $C^{(n)}_i$ in such superpositions behave largely as independent random variables. They are, however, subject to the normalization condition  $\sum_i|C_i^{(n)}|^2=\sum_n|C_i^{(n)}|^2=1$. Also,
the variance of $C^{(n)}_i$ displays a systematic variation with the energy
of the eigenstates and basis states \cite{Bohr:69,Zel}:
\begin{equation}\label{sc0}
\overline{\bigl|C^{(n)}_i\bigr|^2} =\frac{D}{2\pi}\,
\frac{\Gspr}{(E_n-E_i)^2+\Gspr^2/4}\,.
\end{equation}
Here $D$ is the mean level spacing between the basis states (or eigenstates) with a given total angular momentum and parity $J^p$, and
$\Gspr = 2 \pi \overline{H^{2}_{ik}}/D$ is the \textit{spreading width}.
It is determined by the size of the off-diagonal matrix elements of the Hamiltonian $H_{ik}$ which mix the basis states.

Many-body quantum chaos occurs in excited states of all medium and heavy nuclei \cite{Bohr:69,Zel}. It is also typical in atoms and ions with open
$f$ shells. In particular, their excitation spectra demonstrate characteristic Wigner-Dyson level spacing statistics, and the statistics of electromagnetic transition amplitudes is close to Gaussian, which are both signatures of quantum chaos \cite{Ce,Gribakin99,Flambaum02,Gribakin03}.

``Exact'' calculations of the chaotic eigenstates (compound states)
are impossible in principle, since all minor perturbations (e.g., higher-order correlations and relativistic effects) are enhanced due to exponentially small level spacings, and completely 
change the eigenstates. In this case, however, one can use \textit{statistical theory} to predicts physical quantities averaged over a small energy interval containing many compound states. In the problem of electron recombination with ions like W$^{20+}$ and Au$^{25+}$ such averaging occurs naturally,
and the result of the statistical calculation should match the experimental
observation. Indeed, due to a large number of decay channels, the widths of the compound states are two orders of magnitude greater than the exponentially small spacing between neighbouring compound states. As a result, the cross section at any given energy typically contains contributions of $10^2$ or more individual resonances.

Note that due to the extremely strong configuration mixing, approximate quantum numbers such as the orbital occupation numbers (which define
configurations, e.g., $4f^26s5d$), the number of excited electrons $n_e$, the total orbital angular momentum $L$ and spin $S$, are not defined for the compound states. One can only consider average values and distributions for these parameters. For example,
the dependence of the electron orbital occupation numbers on the orbital energy $\varepsilon$ in the chaotic compound states of atoms and ions is close to the Fermi-Dirac distribution $n(\varepsilon,\mu,T)$ where the chemical potential $\mu(E)$ and the effective temperature $T(E)$ depend on the total excitation energy $E$ \cite{Ce,Gribakin99,Gribakin03}.

Earlier papers \cite{Flambaum:93,FV} and reviews \cite{FG95,FG2000}
present the development of the statistical theory for the matrix elements between chaotic compound states. This theory enables one to calculate mean values of orbital occupation numbers, squared electromagnetic
amplitudes, electronic and electromagnetic widths and enhancement of weak interactions in chaotic excited states of  nuclei, atoms and multicharged ions \cite{SF82,Ce,Gribakin99,Flambaum02,Gribakin03,Flambaum:93,FV,FG95,FG2000,random,Dzuba12}.

\section{Electron-ion recombination}

\subsection{Chaotic compound resonances and dielectronic doorway states}

In most ions except bare ones the electron-ion recombination rate is
enhanced by dielectronic recombination (DR). In this process the incident electron excites a target electron to form a quasistationary doubly-excited state which then emits a photon, completing the radiative electron
capture \cite{Massey42}. The DR mechanism often dominates over the direct radiative recombination (RR), and has been the subject of intense theoretical
and experimental work. Experimentally, much progress has been due to the
use of ion storage rings and electron-beam ion traps (EBIT)\cite{Muller9908,Zou03,Beiersdorfer08}.
On the theory side, a number of computational approaches have been used successfully to describe DR for many simpler ions and to produce data for plasma modelling (see \cite{Nahar94,Hahn97,Lindroth01,Tokman02,Badnell03,Behar04,Gu08,Nikolic09,Ballance10,Badnell11} and references therein).

For more complex targets such as U$^{28+}$ or W$^{20+}$, conventional DR approaches severely underestimate measured recombination rates \cite{Mitnik98,Badnell12}. Experiment shows that the recombination rates
at low ($\sim 1$~eV) electron energies in these ions and in Au$^{25+}$ exceed the direct RR rates by two orders of magnitude. At the same time the rates do not show the sharp resonance structure normally associated with DR \cite{Uwira96,HUSF98,schippers11}. Ref. \cite{Gribakin99} explained this phenomenon as being due to electron capture in \textit{multiply} excited, strongly mixed, chaotic eigenstates. It is caused by
the open-shell structure of the compound ion Au$^{24+}$ and the electronic orbital spectrum with no large gaps. These features lead to a very dense spectrum of multiply excited states, as described in Sec.~\ref{sec:chaos}.
The subsequent calculation \cite{Flambaum02} based on the statistical theory provided a quantitative explanation of the enhanced recombination rates near threshold.

Experimentally, direct evidence of \textit{trielectronic} recombination (i.e., via resonances with three excited electrons)  was obtained for Be-like ions (N$^{3+}$, O$^{4+}$, Cl$^{13+}$) \cite{SGBB03,Fogle05}. In these
systems electron capture into a Rydberg state was accompanied by simultaneous $2s^2\rightarrow 2p^2$ excitations. Additionally,
trielectronic and quadruelectronic recombination was observed in Li-like to N-like ions of Ar, Fe and Kr \cite{Beilmann09,Beilmann11,Beilmann13}. It involved intershell excitations leading to $1s^{-1}2p^3$ and $1s^{-1}2p^4$
resonances. However, in the case of chaotic compound resonances
one cannot separate dielectronic, trielectronic or any other
capture process into a resonance with a fixed number of excited electrons. Indeed, a compound state is a chaotic mixture of the states with two, three, four and even five excited electrons, and contributions from all of these configurations are mixed and interfere in the capture amplitude.

Nevertheless, dielectronic states
play a special role. To start with, consider the temporal picture of
radiative recombination. In the first step, the incident electron collides with an ion and excites one electron (by exchanging a virtual photon) and
produces an intermediate state with two excited electrons. We call such state a \textit{doorway state}. This is followed by a ``chain reaction'' in which one of the excited electrons collides with ground state electrons and
excites them. This process continues until all energy of the incident electron equilibrates through excitation of as many electrons as possible.
Thus, the doorway state (with two excited electrons) ``decays'' into other configurations with more excited electrons. This fast internal decay
on timescales $\tau \sim \hbar /\Gspr$, is characterized by the spreading width $\Gspr$ [see Eq. (\ref{sc0})], 
which is several orders of magnitude greater than the autoionization or
radiative widths of the dielectronic state. The notion of the spreading width
is somewhat similar to the quasiparticle width in condensed matter systems
where quasiparticles also decay into other internal excitations of the
system.

Due to the time-energy uncertainty relation the temporal picture cannot be used if the energy of the system is fixed. In this case the configuration
mixing picture is more appropriate. According to this, the dielectronic doorway states are present as components in every chaotic compound state,
and their weights (\ref{sc0}) determine the probability of electron capture into the compound state.
The autoionization width of a compound state $n$ is
$\Gamma^{(a)}_n= \sum_d \Gamma^{(a)}_d|C_d^{(n)}|^2$, where the sum is taken over the dielectronic doorway states whose autoionization widths
$\Gamma^{(a)}_d$ are calculated at the incident electron energy $\varepsilon$.
By the normalization condition $\sum_n|C_d^{(n)}|^2=1$, the sum of the autoionization widths of the compound resonances is equal to the sum of the autoionization widths of the doorway states. Therefore, the energy-averaged total resonance  cross section  may be approximately  described by treating the dielectronic resonances as quasistationary states with the width  $\Gspr$ (see Sec.~\ref{subsec:theory}).

\subsection{Fluorescence yield}

The capture or re-emission of the electron is mediated by the dielectronic doorway states. However, they are not sufficient for describing the process of radiative capture, since a photon can be emitted at any stage of the ``chain reaction''. Three-electron, four-electron and five-electron excited states also radiate, and their total weight in a compound state is several orders of magnitude greater than that of the dielectronic states. As a result, the radiative width $\Gamma^{(r)}_{n}$ of the compound state is enhanced relatively to its autoionization width (since electron emission happens directly from dielectronic states only). As a result, compound states display strongly enhanced fluorescence yields $\omega_f=\Gamma^{(r)}_{n}/\Gamma_{n}$, where 
$\Gamma_{n}=\sum_f\Gamma^{(a)}_{n\to f}+\Gamma^{(r)}_{n}$ is the total width of the compound resonance $n$, and the sum is over all autoionization channels (i.e., autoionization to the ground and  excited states of the ion). 

In Refs. \cite{Flambaum02,Dzuba12} we argued that the fluorescence yield at low incident electron energies is close to 100\%. Indeed, near threshold
only one autoionization channel (with decay to the ground state) is
available, making the autoionization width much smaller than the total radiative width which includes thousands of open photoemission channels. In this case it is sufficient to calculate the total resonant capture cross section to describe recombination. However, at higher electron energies, hundreds of autoionization channels are open (since there are many low-lying excited states in ions with an open $4f$ shell), and the calculation of the fluorescence yield becomes necessary. In the present work we show how this can be done within the statistical theory.

\section{Calculations}

\subsection{Theory}\label{subsec:theory}

The energy-averaged total cross section for electron recombination through the compound resonances is (see, e.g., \cite{Dzuba12})
\begin{equation}\label{aDRcs-final}
\overline{\sigma }_r=\frac{\pi^2}{k^2}\sum_{J^p}\frac{2J+1}{(2J_i+1)}\rho_{J^p}
\left\langle \frac{ \Gamma^{(r)}_{J^p} \Gamma^{(a)}_{J^p\to i,\veps}}
{\Gamma_{J^p}}\right\rangle ,
\end{equation}
where  $k$ is the wave number of the incident electron,
and $i$ denotes the initial (ground) state of the $N$-electron target ion with angular momentum $J_i$. The sum is over the angular momentum and parity
of the compound states,  $\rho_{J^p}=1/ D_{J^p}$ is the level density
of these states for a given $J^p$ in the excited $(N+1)$-electron ion formed by the electron capture, and  $\langle \dots \rangle $ denotes averaging over an energy interval $\Delta \veps \gg D_{J^p}$.

If we assume that the factor in brackets for the dominant $J^p$
in \eqref{aDRcs-final} is approximately the same, we obtain
\begin{equation}\label{aDRcs-final1}
\overline{\sigma }_r
=\frac{\pi^2}{k^2}\,\frac{ \Gamma^{(r)}_n \Gamma^{(a)}_{n\to i,\veps}}
{(2J_i+1)\Gamma_n}\rho=\overline{\sigma }_c \omega_f \,,
\end{equation}
where $\Gamma^{(r)}_n$, $\Gamma^{(a)}_n$, and $\Gamma_n$ are the average widths of
the compound states at energy $\veps$,  $\rho=\sum_{J^p}(2J+1)\rho_{J^p}$ is the total level density of the compound states (which can be found without constructing the states with definite $J^\pi $ from
the Hartree-Fock determinant states), $\omega_f=\Gamma^{(r)}_n/\Gamma_n$ is the average fluorescence yield, and 
\begin{equation}\label{aDRcs-final3}
\overline{\sigma} _c
= \frac{\pi^2}{k^2}\,\frac{\Gamma^{(a)}_{n\to i,\veps}}{(2J_i+1)}\rho\,
\end{equation}
is the energy-averaged total cross section for electron capture into the compound resonances. It is given explicitly by the sum over the dielectric doorways (see \cite{Flambaum02}):
\begin{widetext}
\begin{multline}\label{eq:capture-explicit}
\overline{\sigma }_c
=\frac{\pi^2}{k^2}\sum _{abh ,lj} \frac{\Gamma _{\rm spr}}
{(E_{c_i}+\veps -E_{c_i\to\bar h,a,b})^2 +\Gamma _{\rm spr}^2/4}
\sum _\lambda \frac{\langle a ,b \| V_\lambda \| h ,\veps lj \rangle }
{2\lambda +1} \Biggl[ \langle a ,b \| \hat V _\lambda\| h ,
\veps lj\rangle  - (2\lambda +1) \\
\times \sum _{\lambda '} (-1)^{\lambda +\lambda '+1}\left\{ {\lambda \atop
\lambda '}{j_a \atop j_b }{j \atop j_h }\right\}
\langle b ,a \| \hat V_{\lambda '}\| h,\veps lj\rangle \Biggr]
\frac{n_h }{2j_h +1}\left(1-\frac{n_a }{2j_a
+1}\right)\left(1-\frac{n_b}{2j_b  +1}\right)\,.
\end{multline}
\end{widetext}
Here $E_{c_i}$ is the energy of the ground state target ion with configuration
$c_i$ and $E_{c_i\to \bar h,a,b}$ is the energy of the dielectronic doorway obtained from $c_i$ by making a hole in orbital $h$ and adding electrons in orbitals $a$ and $b$; $E_{c_i}+\veps -E_{c_i\to\bar h,a,b}\approx \veps+\veps_h-\veps_a-\veps_b +Q$, where $\veps_a $, $\veps _b $, and $\veps _h $ are the orbital energies, and $Q$ is the difference in the Coulomb interaction energies (see, e.g., \cite{Gribakin99}); and $n_a$, $n_b$ and $n_h$ are the average occupation numbers of the corresponding orbitals in $c_i$. The two terms in square brackets in Eq.~(\ref{eq:capture-explicit}) represent the direct and exchange contributions,
\begin{align}\notag
&\langle a ,b \| V _\lambda \| h ,c \rangle =
\sqrt{[j_a][j_b][j_h][j_c]}
\xi(l_a +l_c +\lambda )\\ \notag
&\times \xi(l_b +l_h +\lambda )
\left(\!{\lambda \atop 0}{j_a \atop -\frac{1}{2}} {j_c \atop
\frac{1}{2}}\!\right)
\!\!\left(\! {\lambda \atop 0} {j_b \atop -\frac{1}{2} }
{j_h \atop \frac{1}{2}}\!\right)
R_\lambda (a ,b ;h ,c ),
\end{align}
is the reduced Coulomb matrix element, in which $\xi (L)=[1+(-1)^L]/2$ is the parity selection factor, $[j_a]\equiv 2j_a+1$, and
\begin{align}\notag
R_\lambda (a ,b ;h ,c ) &= \iint \frac{r_<^\lambda }
{r_>^{\lambda +1}}\left[f_a (r)f_c (r)+g_a (r)g_c (r)\right]\\ \notag
&\times \left[f_b (r')f_h (r')+g_b (r')g_h (r')\right]\,drdr'\,,
\end{align}
is the radial Coulomb integral, $f$ and $g$ being the upper and lower components of the relativistic radial spinors.

The form of Eq.~(\ref {eq:capture-explicit}) is similar to the expressions which emerge in the
so-called average-configuration approximation \cite{Pindzola86}. The difference between the two approaches is that in a system with chaotic eigenstates, the averaging that leads to Eq.~(\ref {eq:capture-explicit}) occurs naturally due to the strong configuration mixing, rather then being introduced by hand to simplify the calculations.

Note that expressions \eqref{aDRcs-final3} and \eqref{eq:capture-explicit}
allow us to calculate  $\Gamma^{(a)}_{n\to i,\veps}$. In order
to find the fluorescence yield $w_f(\veps)$ we need to calculate the total autoionizing width $\Gamma^{(a)}_{n} = \sum_f \Gamma^{(a)}_{n\to f,\veps}$, where the sum runs over all states of the target ion with energies $E_f< E_i+
\veps$.
The expression for the radiative width of the compound state was given in Ref.~\cite{Flambaum02},
\begin{align}\label{ee1}
\Gamma^{(r)}_d &= \sum_{a,b} \frac{4\omega_{ba}^3}{3c^3}
|\langle a \| d \| b\rangle|^2
\frac{n_b}{2j_b+1}
\left(1 -\frac{n_a}{2j_a+1}\right),
\end{align}
where $n_a$ and $n_b$ are the occupation numbers of orbitals $a$ and $b$ in the compound state at the incident electron energy $\veps$,
$\langle a \| d \| b\rangle$ is the reduced single-electron 
dipole matrix element, and the sum is over $a$ and $b$ such that
$ \omega_{ba}= \veps_b - \veps_a > 0$. The calculations of the spreading
widths give $\Gspr \approx 0.5$~a.u. for Au$^{25+}$ \cite{Gribakin99,Flambaum02,Gribakin03} and 0.68~a.u. for W$^{20+}$ \cite{Dzuba12}.

\subsection{Results}

\begin{figure}[t!]
\includegraphics*[width=85mm]{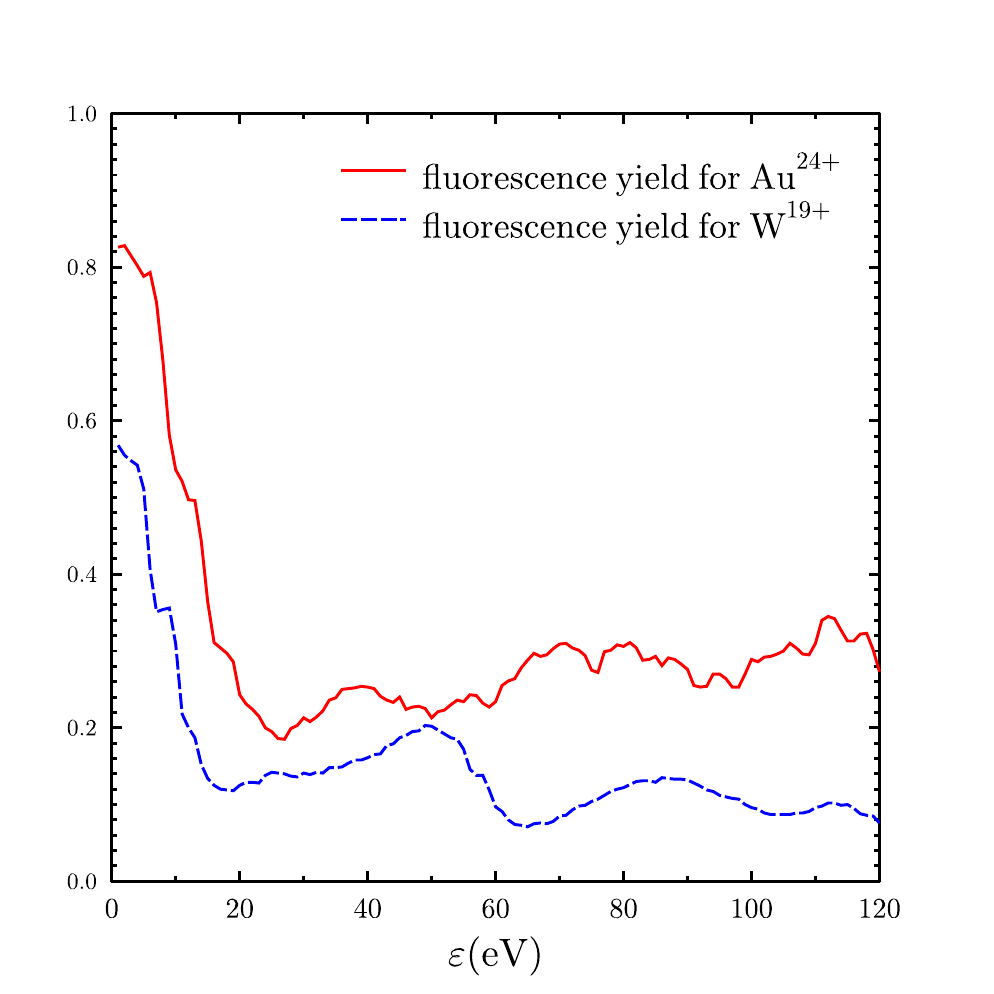}
\caption{(Color online) Calculated fluorescence yields in Au$^{24+}$ and W$^{19+}$ averaged over 1 eV energy interval.}
\label{yield}
\end{figure}

The results of our  calculations of the fluorescence yield are shown in
Fig.~\ref{yield}. At $\veps \lesssim 1$~eV the fluorescence yield in the compound states of Au$^{24+}$ is close to unity, but quickly drops to $\omega _f\sim 0.2$.

\begin{figure*}[ht!]
\includegraphics[width=87mm]{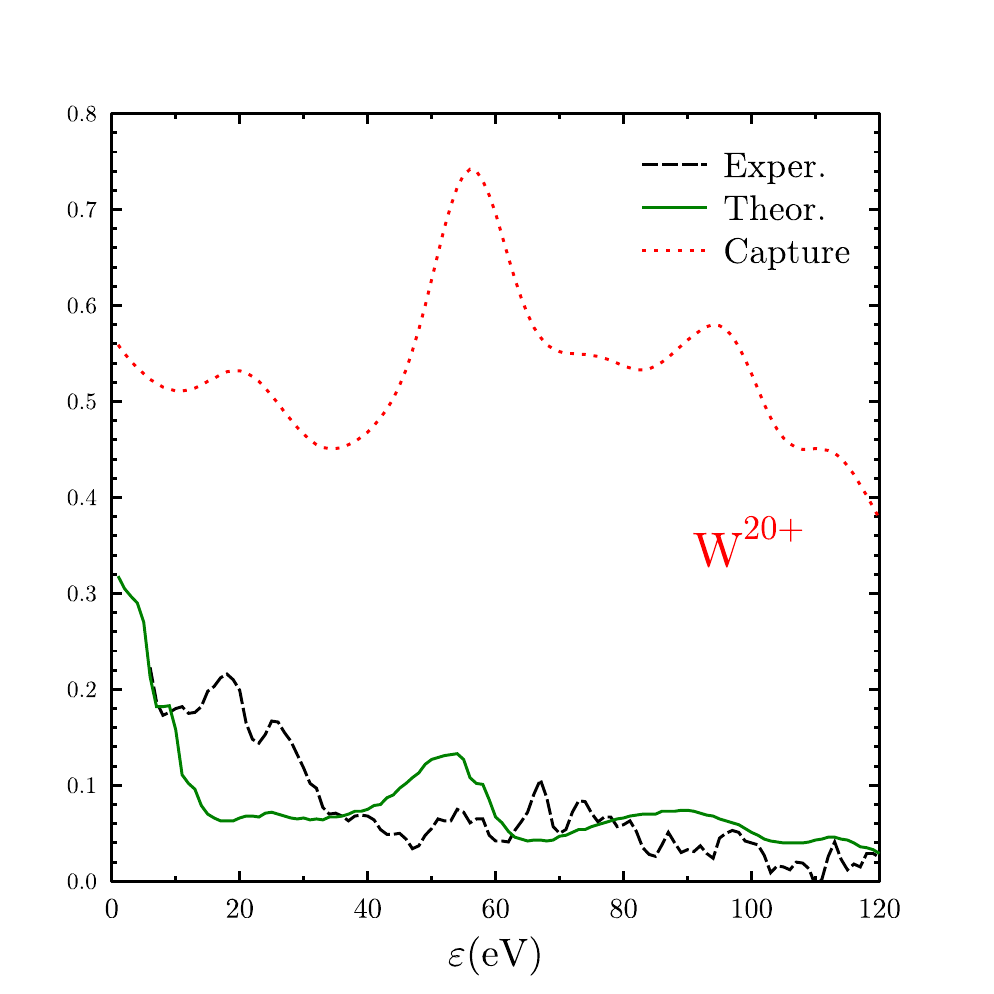}
\includegraphics[width=87mm]{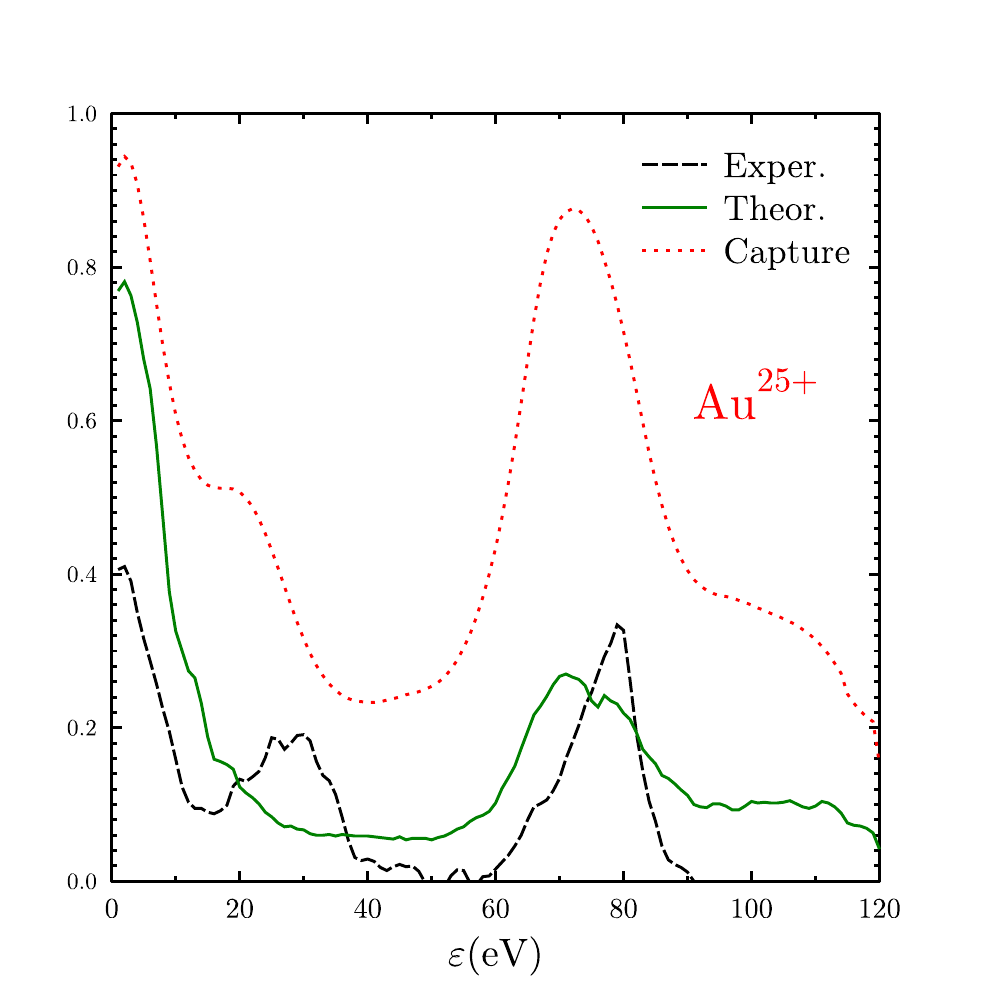}
\caption{(Color online) Reduced cross sections $\sigma k^2/\pi^2$ for electron capture by W$^{20+}$ and Au$^{25+}$, averaged over 1~eV energy intervals to suppress fluctuations. Dotted lines show the calculated total resonant capture ($\overline{\sigma}_c$), solid lines are the calculated recombination cross section $\overline{\sigma}_r=\overline{\sigma}_c \omega_f$, and the dashed lines corresponds to the measured recombination cross section for W$^{20+}$ \cite{schippers11} and Au$^{25+}$ \cite{HUSF98}.}
\label{Au-W}
\end{figure*}

In Fig. \ref{Au-W} the calculated total resonant capture cross section $\overline{\sigma }_c$ and the recombination cross section $\overline{\sigma }_r$ are compared with the experimental data for W$^{20+}$~\cite{schippers11} and Au$^{25+}$~\cite{HUSF98}. To eliminate the strong kinematic dependence the cross sections have been
multiplied by $k^2/\pi^2$. Note that the experimental data display large fluctuations and show some unphysical negative values for the recombination rate (due to background subtraction). To reduce these fluctuations we have averaged the experimental data for $(k^2/\pi^2)\sigma_r$ over 1~eV energy interval. We also averaged over 1 eV range the calculated fluorescence yields (see Fig.~\ref{yield}) to reduce fluctuations in the density of states.

\begin{figure}[htb]
 \includegraphics[width=85mm]{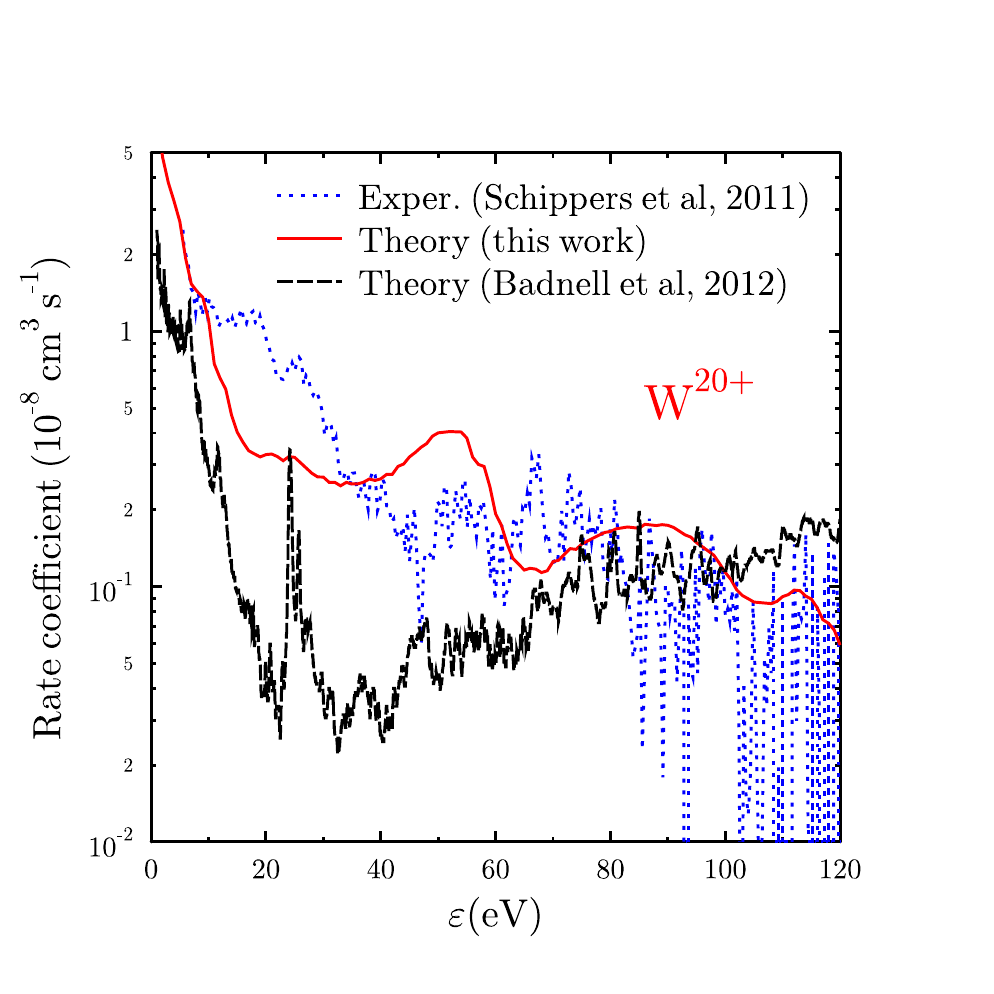}
\caption{(Color online) Recombination rates for  W$^{20+}$. Solid red line is our theory, dashed black line is the calculation of Ref.~\cite{Badnell12}, and dotted blue line is the experimental data \cite{schippers11}.}
\label{W}
\end{figure}

In Fig. \ref{W} we compare the results of our calculations of the recombination rate  for W$^{20+}$ with raw (unaveraged) experimental data \cite{schippers11} and calculations from Ref.~\cite{Badnell12}. 

\subsection{Suppression of photoionization due to chaotic compound resonances}

As discussed above, for energies $\lesssim 1$~eV above the ionization
threshold the radiative width of a compound state can be much greater than its
autoionization width. Besides enhancing recombination, this effect leads to suppression of near-threshold photoionization, since a compound resonance excited by the incident photon will decay primarily by emission of another photon rather than by emitting an electron. Therefore, at these energies the inelastic, Raman photon scattering dominates over the photoionization. Similar to the electron resonant capture, the energy-averaged total photon capture cross section into compound resonances may be approximately described by treating the simple doorway resonances as quasistationary states with the width
$\Gspr$. To obtain the photoionozation cross-sections one should multiply the result by the fluorescence yield calulated in the present work. Taking the corresponding ratios of the widths one can also obtain the cross sections for elastic and inelastic (Raman) photon scattering. 

\section{Conclusions}

We see that in both electron- and photon-induced processes
the interaction between dielectronic and multielectronic configurations
leads to broadening of the dielectronic doorway resonances (due to the internal decay)
and redistribution of the branching between the ``external'' decay
channels in favour of the photoemission.

In principle, within the statistical theory one can take into account the exact quantum numbers of the dielectronic doorway states (angular momentum and parity). One can even diagonalize the Hamiltonian using the basis of the dielectronic states, and then use the ``dielectronic eigenstates'' as the doorway states with the weights from Eq.~(\ref{sc0}).
However, the spreading width $\Gspr$ is comparable to the energy spread of a single configuration. Therefore, our present use of the Hartree-Fock (determinant) basis states as doorway states without a definite $J$
should not significantly reduce the accuracy of the approach. A greater possible error in our calculations is due to the uncertainty in the energies of the doorway states corresponding to the transitions to the ground and excited states of the final $(N+1)$-electron ion, which are  needed to calculate the fluorescence yield. It is likely this uncertainty that leads to a factor-of-two differences between the theoretical and experimental values. However, this discrepancy will be greatly reduced for the Maxwellian, thermally-averaged recombination rates at high electron temperatures.
Such rates are very important in modelling plasmas in astrophysical environments and thermonuclear reactors, and they are less sensitive to the precise doorway state positions. 
A deviation from the experimental data may also be due to the presence of the metastable species in the initial ion beam. Significant contributions of such excited ions was pointed to in the experimental work \cite{schippers11}.

We presented numerical  calculations for W$^{20+}$ and Au$^{25+}$ for which experimental electron recombination data are available \cite{schippers11,HUSF98}. Tungsten is a key plasma-facing component of ITER and future fusion reactors. A broad range of tungsten ions from W$^{20+}$ to W$^{50+}$ is a major plasma impurity and a plasma diagnostic tool. Modelling their fractional abundances and emission spectra reveals that available theoretical recombination rates do not accurately describe the experimental temperature dependence \cite{Putterich08}, and empirical adjustments to the recombination rates were needed to reconcile with the measurements. Clearly, further
experimental and theoretical work on those complex systems is required,
including the extension of our statistical theory calculations to other tungsten ions.

\acknowledgments This work is partly supported by the Australian Research Council and  Russian Foundation for
Basic Research Grants No.\ 11-02-00943. We thank S. Schippers and  A. M\"{u}ller  for providing experimental data in numerical form.



\end{document}